\documentclass{iopart}
\usepackage{iopams,graphicx,hyperref,accents,booktabs}
\usepackage[usenames,dvipsnames]{color}
\usepackage[normalem]{ulem}
\usepackage{subcaption, hhline}

\newtheorem{theorem}{Theorem}
\newtheorem{remark}{Remark}

\begin{document}

\title[On the Hoop conjecture and the weak cosmic censorship conjecture]{On the Hoop conjecture and the weak cosmic censorship conjecture for the axisymmetric Einstein-Vlasov system}
\author{Ellery Ames$^{1}$, H\r{a}kan Andr\'easson$^{2}$ and Oliver Rinne$^{3}$}
\address{ 
  $^{1}$ Department of Mathematics, Humboldt State University, 1 Harpst St., Arcata, California 95521, USA\\ 
  $^{2}$ Mathematical Sciences, Chalmers University of Technology and University of Gothenburg, S-41296 Gothenburg, Sweden\\
  $^{3}$ Hochschule f\"ur Technik und Wirtschaft Berlin, Fachbereich 4, Treskowallee 8, 10318 Berlin, Germany}
\ead{ellery.ames@humboldt.edu, hand@chalmers.se, oliver.rinne@htw-berlin.de}

\begin{abstract}
We consider gravitational collapse for the axially symmetric Einstein-Vlasov system. We investigate the weak cosmic censorship conjecture in the case of highly prolate initial data and we investigate the ``only if" part of the Hoop conjecture. Shapiro and Teukolsky initiated a similar study in 1991 \cite{Shapiro1991} where they found support that the weak cosmic censorship conjecture was violated for sufficiently prolate spheroidal initial data. More recently, independent studies of this problem have been carried out by Yoo, Harada and Okawa \cite{Yoo2017} and by East \cite{East2019}. 
A common feature in these works is that the initial data are dust-like. 
Dust can be considered as a \textit{singular} case of matter described by the Einstein-Vlasov system. 
The original motivation by Shapiro and Teukolsky to study this problem is based on the Lin-Mestel-Shu instability for gravitational collapse of uniform spheroids in the case of dust in Newtonian gravity. We argue that the Lin-Mestel-Shu solution is not relevant for studying the weak cosmic censorship of the Einstein-Vlasov system and we argue that dust-like initial data is also not relevant. To investigate collapse of highly prolate spheroidal configurations for the Einstein-Vlasov system is nevertheless interesting in view of the Hoop conjecture. By choosing highly prolate initial data the weak cosmic censorship conjecture is seriously tested. We carry out such a study for initial data which are not dust-like. We find formation of an apparent horizon in all cases we consider, which provides support for the weak cosmic censorship conjecture. In our tests of the Hoop conjecture we compute the polar circumference $\mathcal{C}_{H,p}$ at the time when the apparent horizon forms and find that it is less than 12\% above $4\pi M$, where $M$ is the irreducible mass of the apparent horizon, which agrees with the spirit of the Hoop conjecture. 
 
\end{abstract}
 
\pacs{
  04.20-q,  %Einstein equations 
  04.40.-b, %self-gravitating systems
  52.65.Ff, %Vlasov equation.
  04.25.dc %numerical relativistic studies of critical behavior, singularities, and cosmic censorship.
}

%%%%%%%%%%%%%%%%%%%%%%%%%%%%%%%%%%%%%%%%%%%%%%%%%%%%%%%%%%%%%%%%%%%%%%%%%
%%%%%%%%%%%%%%%%%%%%%%%%%%%%%%%%%%%%%%%%%%%%%%%%%%%%%%%%%%%%%%%%%%%%%%%%%
%%%%%%%%%%%%%%%%%%%%%%%%%%%%%%%%%%%%%%%%%%%%%%%%%%%%%%%%%%%%%%%%%%%%%%%%%

\section{Introduction}
\label{s:introduction}
In 1991 Shapiro and Teukolsky \cite{Shapiro1991} considered the axially symmetric Einstein-Vlasov system and studied gravitational collapse of prolate spheroidal configurations. They found numerical support that the weak cosmic censorship conjecture was violated in the case of highly prolate spheroidal configurations. The motivation for posing such data is related to the Hoop conjecture which states that an imploding object forms a black hole when, and only when, a circular hoop with a specific critical circumference could be placed around the object and rotated about its diameter. Hence, according to the Hoop conjecture, if a sufficiently elongated body undergoes collapse then no apparent horizon will form and a naked singularity will result. This was studied analytically by Thorne in the case of cylindrical symmetry \cite{Thorne1972}, i.e. when the body has infinite extent. 

Criticism about the work \cite{Shapiro1991} was raised by Rendall \cite{Rendall1992} since Shapiro and Teukolsky do not consider the (regular) Einstein-Vlasov system but rather the Einstein-dust system. The relation between dust and Vlasov matter can be explained as follows, cf. \cite{Rendall1992}. 
The unknown in the Vlasov equation is the phase space density function $f$. The Vlasov equation is linear in $f$ and distributional solutions therefore make sense. One class of distributional solutions is given by
\[
f(x^{\gamma},p^a)=-u_0|g|^{-1/2}\rho(x^{\gamma})\delta(p^a-u^a),
\]
where $\rho\geq 0$ and $u^a(x^{\gamma})$ is a mapping from spacetime into the mass shell and $u_0$ is given by $u^a$ from the mass shell relation.
Solutions of the Einstein-Vlasov system where
the phase space density $f$ has this form are in one-to-one correspondence with dust solutions
of the Einstein equations with density $\rho$ and four-velocity $u^{\alpha}$.
Dust may thus be considered as a \textit{singular} case of matter described by the Vlasov equation. 
This fact is of fundamental importance when singularity formation is investigated for the Einstein-dust system and the Einstein-Vlasov system respectively, as will be discussed in detail in Section 2. 

More recently, the problem considered by Shapiro and Teukolsky in \cite{Shapiro1991} 
has been reconsidered in two independent works; in 2017 by Yoo, Harada and Okawa \cite{Yoo2017} and in 2019 by East \cite{East2019}. Despite the criticism raised by Rendall \cite{Rendall1992}, these studies again consider dust-like initial data as is done in \cite{Shapiro1991}. (The terminology dust-like will be specified below.) 
Of course, by reconsidering the study \cite{Shapiro1991} it is natural to pose the same type of initial data but by neglecting a discussion about the relation between dust and Vlasov matter these studies give a misleading picture of gravitational collapse for the (regular) Einstein-Vlasov system. Indeed, the original motivation by Shapiro and Teukolsky for studying gravitational collapse of highly prolate spheroidal initial data is related to the Lin-Mestel-Shu instability \cite{Lin1965}. This instability occurs when prolate spheroids of dust undergo gravitational collapse in Newtonian gravity. In Section \ref{S2} the relation between dust and Vlasov matter in Newtonian gravity is discussed in detail and we then argue in Section \ref{S3} that the Lin-Mestel-Shu instability is not relevant as starting point for studying gravitational collapse of the Einstein-Vlasov system. The relation between these matter models in the framework of general relativity is outlined in Section \ref{S3} and \ref{S4}. In particular we discuss the phenomenon of critical collapse. This is a distinguishing feature between these matter models; critical collapse only occurs for Vlasov matter. This crucial difference makes the study of gravitational collapse for these two matter models fundamentally different. In the present work we investigate collapse of highly prolate configurations as in \cite{Shapiro1991,Yoo2017,East2019} but for the (regular) Einstein-Vlasov system. 
In view of the Hoop conjecture this choice of initial data constitutes a serious test of the weak cosmic censorship conjecture. We find no sign of singularity formation before an apparent horizon forms which provides strong support for weak cosmic censorship. By the time the apparent horizon forms the matter configuration has changed drastically and the initial highly prolate shape is ``washed out". We also investigate the geometry of the apparent horizon and we find that it is mildly prolate, i.e. the polar circumference is about 15\% larger than the equatorial circumference. We then investigate the Hoop conjecture, or more precisely the ``only if" part of the conjecture, cf. Section \ref{S45}. We choose to interpret the conjecture in terms of the geometry of the apparent horizon, cf. the discussion in Section \ref{S45}. A similar interpretation is made in \cite{East2019}.  
We find that
\[
\frac{\mathcal{C}_{H,p}}{4\pi M_H}<1.12\; \textrm{ and }\; \frac{\mathcal{C}_{H,e}}{4\pi M_H}<1,
\]
at the time when the horizon forms. Here $\mathcal{C}_{H,p}$ and $\mathcal{C}_{H,e}$ are the polar and equatorial circumference respectively and $M_H$ is the horizon mass. This is in line with the spirit of the Hoop conjecture. 

The outline of the paper is as follows. In Section \ref{S2} the relation between dust and Vlasov matter is discussed in Newtonian gravity where rigorous results are available. In Section \ref{S3} the relation between these matter models is discussed in the relativistic setting, and in Section \ref{S4} we turn to the phenomenon of critical collapse. 
This phenomenon only occurs for Vlasov matter and not for dust. 
We discuss our interpretation of the ``only if" part of the Hoop conjecture in section \ref{S45} where we also give the details of the geometric quantities involved. The Einstein-Vlasov system and the numerical method are briefly discussed in Section \ref{S5}. The type of initial data we investigate in our simulations is described in Section \ref{S6}. In Section \ref{S8} we finally discuss the numerical results of our investigation. A test of the numerical accuracy of our code is presented in \ref{numerical_accuracy}. 

\section{The relation between dust and Vlasov matter in Newtonian gravity}\label{S2}
It is useful to compare dust and Vlasov matter first in the case of Newtonian gravity, since there are a number of rigorous results available.
In Newtonian gravity a comparison between dust and Vlasov matter means a comparison between solutions of the pressureless Euler-Poisson system and solutions of the Vlasov-Poisson system. 

Rein and Taegert \cite{ReinTaegert} have investigated the relation between these systems. Let us summarize their result. The pressureless Euler-Poisson system reads 
\begin{eqnarray*}
& &\partial_t\rho+\nabla\cdot(\rho u)=0,\\
& &\partial_t u+(u\cdot\partial_x)u=-\partial_x U(t,x),\\
& & \Delta U=4\pi\rho, \;\; \lim_{|x|\to\infty}U(t,x)=0,
\end{eqnarray*}
and the Vlasov-Possion system reads
\begin{eqnarray*}
& &\partial_t f+v\cdot\partial_x f-\partial_x U\cdot\partial_v f=0,\\
& & \Delta U=4\pi\rho, \;\; \lim_{|x|\to\infty}U(t,x)=0,\\
& &\rho(t,x)=\int f(t,x,v)\, dv.
\end{eqnarray*}
Here $u$ is the velocity field, $\rho$ the density, $U$ the potential and $f$ is the density function on phase space. 
Let
\begin{equation}\label{Dindata}
\rho(t,x):=\frac{3}{4\pi}\frac{1}{r^3(t)}\mbox{\textbf{1}}_{B_{r(t)}(0)},
t\geq 0, x\in \mathbb{R}^3,
\end{equation}
where $r(t)$ solves
\[
\ddot{r}=-\frac{1}{r^2},\; r(0)=1,\; \dot{r}=0,
\]
and where $B_r(0)$ denotes the ball centered at the origin of radius $r$. 
Also, let
\[
u(t,x)=\frac{\dot{r}(t)}{r(t)}x,
\]
then $(\rho, u, U)$ is a solution of the Euler-Poisson system above (where $U$ is determined via the Poisson equation). 
This solution describes a ball of dust, initially at rest, which collapses under its own gravitational field to a point in finite time since it can be shown that $\lim_{t\to T}r(t)=0$ for some $T>0$. In fact, the initial density, which above has amplitude $3/(4\pi)$ so that the total mass is one, can be chosen to have an arbitrarily small amplitude without changing the conclusion; it only affects the collapse time $T$ which will be larger with a smaller amplitude. 
If we swap matter model from dust to Vlasov, then the global existence result for the Vlasov-Poisson system \cite{Pfaffelmoser,LionsPerthame} guarantees that no singularity will form. 
The global existence result says nothing about the behaviour of the solutions, only that they will not break down. In the work by Rein and Taegert the following result is shown. 
\begin{theorem}[\cite{ReinTaegert}]
For any constants $C_1,C_2 > 0$ there exists a smooth, spherically symmetric solution $f$ of the Vlasov-Poisson system such that initially
\[
\|\rho(0)\|_{\infty}<C_1,
\]
but at some time $t^*>0$
\[
\|\rho(t^*)\|_{\infty}>C_2.
\]
\end{theorem}
Hence an arbitrarily small initial density can be prescribed such that at some time $t^*$ the density becomes arbitrarily large in the evolution. Such a solution thus approximates the behaviour of the solution of the Euler-Poisson system with the important difference that only the latter forms a singularity. 
 
A simple picture to understand why dust solutions form singularities whereas Vlasov solutions do not is as follows. In the case of dust let us think of particles uniformly placed on a sphere at rest at time zero and then evolved by the induced self-gravity. The particles reach the centre at the same time causing the density to become unbounded and the solution thus blows up. Corresponding initial data for Vlasov matter mean that the particles are not perfectly at rest initially, there is some dispersion in phase space. Hence they will not reach the centre exactly at the same time. Although the density becomes large the solution does not blow up which is ensured by the global existence results mentioned above. 

It is instructive to specify in mathematical terms the relation between the initial data $\mathring{\rho}$ of the Euler-Poisson system and the initial data $\mathring{f}$ of the Vlasov-Poisson system in the situation where the aim is that the solutions of the latter system should approximate solutions of the former system. The initial data are related as follows: 
\begin{equation}\label{DVindata2}
\mathring{f}(x,v)=h_{\epsilon}(v)\mathring{\rho}(x),
\end{equation}
where $h_{\epsilon}$ tends to the Dirac delta function as $\epsilon\to 0$. 
In the example above, the initial data ({\ref{Dindata}) for the Euler-Poisson system is chosen 
such that $\mathring{\rho}(x)=c 1_{[0,1]}(|x|)$ with $c=3/(4\pi)$. 
For such a choice, the initial data for the Vlasov-Poisson system takes the form 
\begin{equation}\label{DVindata}
\mathring{f}(x,v)=c\, h_{\epsilon}(v)1_{[0,1]}(|x|).
\end{equation}
Hence, for a uniform density there are two constants involved to describe the initial data $\mathring{f}$; $\epsilon$ and $c$. The parameter $\epsilon$ determines how closely the solution to the Vlasov-Poisson system approximates solutions of the Euler-Poisson system and $c$ determines the amplitude of the macroscopic density $\rho$. Note that for a given value of $c$ the amplitude of the phase space density $\mathring{f}$ tends to infinity when $\epsilon\to 0$, i.e. the amplitude of the macroscopic density $\mathring{\rho}$ and of the phase space density $\mathring{f}$ can be very different. This is an important observation for the discussion in the following sections. 

\begin{remark}
In order to guarantee local existence of solutions the initial data $\mathring{\rho}$ in (\ref{DVindata}) should in fact be continuously differentiable. In case the data for dust is as in the example above, which is discontinuous at $r=1$, it is natural to choose a mollified version of such data for the Vlasov-Poisson system, cf. \cite{ReinTaegert}. For the present discussion this technicality is however not essential. 
\end{remark}

\section{The relation between dust and Vlasov matter in general relativity}\label{S3}
The discussion above concerned singularities in the Newtonian situation. In the general relativistic framework singularities form for both the Einstein-Vlasov system and the Einstein-dust system \cite{Christodoulou1984,AndreassonKR2011,Andreasson2012} but there is a fundamental difference also in this case. Namely, an initial spherically symmetric uniform density $\mathring{\rho}(x)=c 1_{[0,1]}(|x|)$ for the Einstein-dust system will evolve and form a black hole independently of the size of $c$. The time until the black hole forms depends on $c$ but a black hole forms for any $c>0$. For the Einstein-Vlasov system on the other hand, if the initial amplitude of the phase space density $\mathring{f}$ is sufficiently small, then global existence holds, the spacetime is geodesically complete and no black holes form, cf. \cite{Rein1992} for the spherically symmetric case and \cite{Lindblad2019,Fajman2021} for the general case. In this case the fields are weak, the relativistic effects are small and the solutions are close to solutions of the Vlasov-Poisson system, cf. \cite{ReinRe1992} where the Newtonian limit of solutions to the Einstein-Vlasov system is studied. 

The discussion above is essential for the present topic about collapse of prolate spheroidal configurations. As mentioned in the introduction, the original motivation of Shapiro and Teukolsky for considering this topic is related to the Lin-Mestel-Shu instability \cite{Lin1965} for collapse of prolate and oblate spheroidal configurations of dust in Newtonian gravity. Let us focus on the prolate case in this discussion. It is shown in \cite{Lin1965} that a prolate spheroid of dust collapses to a spindle so that the density becomes unbounded similarly to the situation when a uniform ball of dust collapses to a point. From the global existence results for the Vlasov-Poisson system we can again conclude that the scenario is different in the case of Vlasov matter; the density will become large but it will stay bounded. Now, the initial data that Shapiro and Teukolsky considered \cite{Shapiro1991} and which in the evolution indicated violation of the weak cosmic censorship conjecture, is inspired by the initial data for the Lin-Mestel-Shu instability in the prolate case. Such initial data can be described by the parameters $a,b$ and $M$, where $a$ is the equatorial radius, $b$ is the semi-major axis and $M$ is the mass. Shapiro and Teukolsky consider a family of initial data where $a<b$, with fixed eccentricity $e=\sqrt{1-a^2/b^2}=0.9$, and with varying ratio $b/M$. The ratio $b/M$ determines how compact the configuration is. The relativistic effects are stronger in the case of a compact body with a small ratio $b/M$, whereas a body with a large ratio is close to being Newtonian. Shapiro and Teukolsky consider two cases for this ratio: $b/M=2$ and $b/M=10$. In the former case, which describes a compact configuration, they find formation of a trapped surface and collapse to a black hole. It is the latter case which indicates violation of cosmic censorship in \cite{Shapiro1991}. However, in this case the ratio $b/M$ is large 
which means that spacetime resembles the Newtonian case. Hence, if one chooses dust-like initial data in the sense that $h_{\epsilon}$ is replaced by the Dirac delta function in (\ref{DVindata2}), then the solution will be close to the dust solution in the Newtonian case. This solution is the Lin-Mestel-Shu solution which develops a singularity. Shapiro and Teukolsky do choose dust-like initial data and find that their solutions resemble the Lin-Mestel-Shu solutions which is thus not surprising. Since the situation is very different for Vlasov matter, in the sense that solutions of the Vlasov-Poisson system \textit{do not form} singularities, it is highly unlikely that solutions of the Einstein-Vlasov system in the Newtonian regime will develop singularities. 

\begin{remark}
We use the terminology \textbf{dust-like} when we refer to the initial data, and to the solutions, in  \cite{Shapiro1991,Yoo2017,East2019}. The reason is that the initial data for the phase space density $\mathring{f}$ is not explicitly given so it is difficult to know the exact form of the initial data. The authors write that initially the particles are at rest and have no angular momentum which indicate that they take $\epsilon=0$ in (\ref{DVindata}). Moreover, in \cite[p.5]{East2019} the author writes that the density blows up due to shell crossing. For the (regular) Einstein-Vlasov system shell crossing cannot occur, cf. \cite{Rendall1996,Andreasson2011}. 
However, even if $\epsilon=0$ the numerical errors will induce some dispersion in the momentum variables in the evolution so that the solutions do not exactly describe dust. Hence, by the notion dust-like we have in mind initial data that describe dust but where the solutions for $t>0$ may somewhat resemble regular solutions of the Einstein-Vlasov system although they will be very close to the corresponding dust solutions. 
\end{remark}

A natural question to ask is what the expected behaviour of solutions of the (regular) Einstein-Vlasov system is when the ratio $b/M$ is large. Let us fix $\epsilon>0$ in (\ref{DVindata2}) and let us fix $a$ and $b$, with say eccentricity $e=0.9$, although this is not essential for the following discussion. In order to investigate the evolution of initial data with a large ratio $b/M$ we should choose $M$ small. For simplicity we may think of an amplitude $c$ in front of a given $\mathring{\rho}$, as in (\ref{DVindata}), which we then choose small to ensure that $M$ is small. 
In the limit $M\to 0$ we have $c\to 0$ which implies that the amplitude of $\mathring{f}$ goes to zero since $\epsilon$ is fixed. Now, for a sufficiently small amplitude of $\mathring{f}$, the global existence results for the Einstein-Vlasov system \cite{Lindblad2019,Fajman2021} can be applied which results in a geodesically complete spacetime, i.e., no singularities develop whatsoever. 
Hence, the relevance of the original motivation by Shapiro and Teukolsky for investigating the weak cosmic censorship conjecture for the Einstein-Vlasov system is highly questionable; the regime they investigate to find evidence of violation of cosmic censorship is the Newtonian regime and the solutions they consider are dust-like. In view of the discussion above it is very unlikely that there will be any violation of cosmic censorship in this regime. 

\section{Critical collapse of prolate initial data}\label{S4}
The discussion above leads naturally to the topic of critical collapse. In studies of critical collapse, initial data of the form $A\Psi$ is studied where $A>0$ is a positive constant and $\Psi$ is a given function. If critical collapse occurs there is a critical number $A_*$ such that if $A<A_*$, then the evolved solution is regular (disperses, oscillates or is a steady state) whereas if $A>A_*$, then the solution undergoes gravitational collapse which results in black hole formation if the weak cosmic censorship conjecture holds true. We refer to these two situations as the subcritical case and the supercritical case respectively. It follows from the discussion above that critical collapse \textit{does not occur} for the Einstein-dust system since by taking initial data as
\begin{equation}\label{dustin}
\mathring{\rho}=A\, 1_{[0,1]}(|x|),
\end{equation}
black holes form in the evolution for any $A>0$, cf. \cite{OppenheimerSnyder1939}. Hence there is no critical amplitude $A_*$ and hence no critical collapse. For the Einstein-Vlasov system on the other hand, it is well-known that critical collapse occurs, cf. \cite{Olabarrieta2001,Andreasson2006} in the spherically symmetric case and \cite{Ames2021} in the axially symmetric case. This again shows that there are fundamental differences between dust and Vlasov matter although dust can be approximated arbitrarily well with Vlasov matter, cf. \cite{ReinTaegert,Andreasson2023}. 

In the present work we investigate gravitational collapse of highly prolate initial configurations for the (regular) Einstein-Vlasov system. In particular we investigate the weak cosmic censorship conjecture. We find such a study motivated since it is in general an open question whether or not weak cosmic censorship holds true for this system. Serious attempts to find initial data leading to naked singularities have been made in the spherically symmetric case, cf. \cite{RendallVel2017}, but so far without success. In \cite{Ames2021} gravitational collapse (and in particular critical collapse) was investigated for some classes of axially symmetric initial data. The data considered were toroidal and the particles were far from the axis of symmetry initially. One reason for such a choice was due to numerical difficulties related to the axis.
For the present work, certain difficulties related to the axis have been resolved, as discussed in Section \ref{S5}. 
Another reason the case of highly prolate initial data is of interest for investigating the weak cosmic censorship conjecture is due to the Hoop conjecture, cf. \cite{Thorne1972}, which states that an imploding object forms a black hole when, and only when, a circular hoop with a specific critical circumference could be placed around the object and rotated about its diameter. Hence, highly relativistic prolate initial configurations challenge the weak cosmic censorship conjecture; if the Hoop conjecture and the weak cosmic censorship conjecture hold true then the matter configuration has to change shape drastically before collapsing. This is indeed what we find in our simulations, cf. Section \ref{S8}. 

The form of the initial data we study is given in detail in Section \ref{S6} but let us make some remarks. Roughly, we fix $\epsilon>0$ in (\ref{DVindata2}), which means that our initial data are not dust-like. 
A consequence of this is that \textit{if we consider initial data in a (sufficiently) Newtonian regime} then the solutions disperse and no gravitational collapse occurs. It is only in the case of dust that solutions in a Newtonian  regime will undergo gravitational collapse as is the case in \cite{Shapiro1991,Yoo2017,East2019}. 
The initial data considered in \cite{Shapiro1991,Yoo2017,East2019} become more Newtonian the larger the ratio of $b/M$ is. The largest value in \cite{Shapiro1991,Yoo2017} is $10$ whereas it is $20$ in \cite{East2019}. 
Since our aim is to investigate the weak cosmic censorship conjecture we consider supercritical initial data to ensure that collapse occurs.
For the choice of $\epsilon$ we make (or in other words the level of dispersion we impose), an initial datum with $b/M\geq 10$ is subcritical and is not considered here. We could include subcritical data but due to the highly prolate initial configurations our numerical domain is rather limited in the radial direction (in cylindrical coordinates) and since subcritical typically solutions disperse the particles leave the numerical domain quickly and the simulations become fruitless.  

\section{The Hoop conjecture and the dynamics of apparent horizons}\label{S45}
The Hoop conjecture was formulated 1972 by Kip Thorne in \cite{Thorne1972}: 

\begin{quote}
Horizons form when and only when a mass M gets compacted into a region whose circumference in EVERY direction is $\mathcal{C}\leq 4\pi GM/c^2$. (Like most conjectures, this one is sufficiently vague to leave room for many different mathematical formations!)
\end{quote}
This conjecture has been investigated both analytically and numerically in many works since then. Senovilla \cite{Senovilla} has also proposed a reformulation of the conjecture. We refer
to this work and also to \cite{ChoptuikLP2015}, and the references therein, for an overview of previous studies of the conjecture. 

In this work we will only be concerned with the ``only if" part of the conjecture due to its relation to the weak cosmic censorship conjecture but let us mention that the ``if" part has been studied in \cite{ChoptuikP2010} where numerical support for the conjecture was found. 

The ``only if" part of the Hoop conjecture reads: 
If a horizon forms then the mass $M$ is compacted into a region whose circumference $\mathcal{C}$ in every direction is $\mathcal{C}\leq 4\pi M$. 

One can ask the question at which instant of time this inequality should be satisfied. Does it refer to the initial data? Clearly, it does not refer to the initial data since there are known examples that apparent horizons form in the evolution of initial data which do not satisfy this inequality, cf. \cite{Andreasson2012,OppenheimerSnyder1939}. 
How should one interpret the circumference of a body? For Vlasov matter, even if the data is chosen in such a way that there is a clear cut boundary initially it may not be so later on in the evolution. A typical situation is that there is a core of the matter which is surrounded by a thin atmosphere. This is what happens in the simulations in this work. A fraction of the particles are ejected outwards and form a thin atmosphere surrounding the core of the matter which eventually collapses to a black hole. This  atmosphere can extend far out. Hence, it is also not clear what the mass $M$ refers to if one cannot naturally define the boundary of the body. 

In this work we have chosen the following interpretation of the ``only if" part of the conjecture. Assume that there is no apparent horizon initially and that an apparent horizon forms at $t=t_H>0$. The circumference $\mathcal{C}$ in the conjecture is then the polar and the equatorial circumference of the apparent horizon, which we denote by $\mathcal{C}_{H,e}$ and $\mathcal{C}_{H,p}$ respectively. For the mass $M$ we choose the horizon mass $M_H$ which we define to be the irreducible mass $M_{irr}$ of the apparent horizon. Note that for a stationary black hole the irreducible mass $M_{irr}$ equals the mass of the black hole when the total angular momentum vanishes. 
%Since the total angular momentum vanishes in our case we use that the mass of the black hole equals the irreducible mass $M_{irr}$ of the apparent horizon. 
The horizon mass is given by $M_H:=M_{irr}=\sqrt{\mathcal{A}_H/16\pi}$, 
where $\mathcal{A}_H$ is the area of the apparent horizon. 
A similar interpretation of the ``only if" part of the Hoop conjecture is used by East \cite{East2019}. 

Using this interpretation we can now specify in precise terms how we compute the geometric quantities.  As described in \cite{Ames2021}, an apparent horizon finder is implemented in the code. 
Recall that an apparent horizon is the outermost two-surface in a spatial slice whose outgoing null expansion vanishes. Such a surface is a curve in cylindrical coordinates $(r,z)$ since one dimension is suppressed. We parametrize this curve by the spherical polar angle $\theta$ as 
\begin{equation}\label{paraAH}
r=R(\theta)\sin{\theta},\;\;\; z=R(\theta)\cos{\theta},
\end{equation}
where $R$ is the spherical polar radius. 
In view of the form of the spatial metric, \cite[p.5]{Ames2021}, the mathematical expressions for the polar and equatorial circumferences read 
\begin{eqnarray*}
& &\mathcal{C}_{H,p}=4\int_0^{\pi/2}\psi^2 e^{rs}\sqrt{R^2+(R')^2}\,d\theta,\\
& &\mathcal{C}_{H,e}=2\pi\big(\psi^2 e^{rs} R)_{\theta=\pi/2}.
\end{eqnarray*}
Here $R'$ denotes the derivative with respect to $\theta$, and $\psi$ and $s$ are metric fields. 
We also need an expression for $\mathcal{A}_H$ in order to compute the irreducible mass. This is given by 
\[
\mathcal{A}_H=2\pi \int_0^{\pi}\psi^4 e^{rs}  \sqrt{R^2+(R')^2} \, R\sin{\theta}\,d\theta.
\]
From this formula we get the horizon mass $M_H$ as described above. 

The quantities $\mathcal{C}_{H,p}, \mathcal{C}_{H,e}$ and $M_{H}$ that enter in the formulation of the conjecture depend on the time $t$ where $t\geq t_H$. We stop our simulations shortly after an apparent horizon has formed since we then experience increasing violations of the constraints and of the mass conservation. Hence, in Section \ref{S8} where we present the results of our simulations, we compute the quantities
\begin{equation}\label{kappa}
\mathcal{\kappa}_p:=\frac{\mathcal{C}_{H,p}}{4\pi M_H}
\quad \textrm{and} \quad 
\mathcal{\kappa}_e:=\frac{\mathcal{C}_{H,e}}{4\pi M_H},
\end{equation}
only during a short time after the apparent horizon has formed. 
For the prolate data that we consider we find that $\mathcal{\kappa}_p<1.12$ and that $\mathcal{\kappa}_e<0.9$ at $t=t_H$. These bounds persist on the small time interval we consider after the formation of a horizon. Hence our results are in line with the spirit of the Hoop conjecture and they are also in line with the results in \cite{East2019}. 

\section{The Einstein-Vlasov system and the numerical method}\label{S5}
The formulation of the Einstein-Vlasov system that we use is given in Section 2 in \cite{Ames2021} and we refer to this work for the details of the system of equations. The numerical method that the simulations rely on is based on the particle in cell (PIC) method which is described in Section 3 in \cite{Ames2021}, see also \cite{HockneyEastwood1988}. To a large extent we use the same code in the present investigation as in \cite{Ames2021} but with an essential modification. In \cite{Ames2021} the initial data were chosen such that particles initially were far from the axis of symmetry since numerical difficulties arise close to the axis due to the coordinate singularity of the cylindrical coordinates. There are two numerical difficulties related to the axis. One is the problem how to propagate particles when they are close to the axis. The cylindrical radius $r$ of their position is then small and it is necessary that $r$ remains nonnegative also after the particles have been propagated. The second problem is that some terms in the equations contain singular factors of $1/r$, cf. Equations (53) and (54) in \cite{Ames2021}. 
In the present work we have modified the code to treat the first of these difficulties in a more efficient way compared to the method used in \cite{Ames2021} (outlined in Section 3.2.5 in \cite{Ames2021}).  
The modification is that Cartesian coordinates are used to propagate the particles, i.e. we use cylindrical coordinates as in \cite{Ames2021} everywhere in the code except for the part where the particles are evolved. More precisely, when the code reaches the point where the particles are going to be propagated, we transform to Cartesian coordinates and evolve the particles for one time step, and then transform back to cylindrical coordinates. In this way the axis of symmetry causes no numerical problems for the particle propagation and the loss in the speed of computation, due to the increase of the dimension by using Cartesian coordinates instead of cylindrical coordinates, is reduced to a minimum. Presently we have only been able to implement this strategy in the case of vanishing total angular momentum. This is related to the fact that certain components of the metric do not appear directly in our $(2+1)+1$ formalism, but only in form of a twist vector, which contains certain combinations of first derivatives of those metric components. Hence we have restricted the study to such initial data but point out that the same restriction is present in \cite{Shapiro1991,Yoo2017,East2019}. For future investigations it would be valuable to implement this strategy also in the general case. Let us here mention that the work \cite{Abrahams1994} extended the study \cite{Shapiro1991} to also include rotating spacetimes.

\section{The initial data}\label{S6}
We investigate a family of highly prolate initial data which is sufficiently compact to collapse and we track the formation of an apparent horizon.  
The details of the initial data that we use in the simulations are given below. For the notation we refer to \cite{Ames2021} except that we use a slightly different choice of momentum variables. Namely, the momentum variable $v_3$ in \cite{Ames2021} is replaced by the variable $b:=r v_3$ so that $f=f(t,r,z,v_1,v_2,b)$. In fact, the density function that we evolve is the rescaled density function $\hat f= \psi^8 e^{2rs} f,$ cf. Appendix A.1 in \cite{Ames2021}. We anyhow denote the initial data for $\hat f$ by $\mathring{f}$. The family of initial data used in the simulations is given by 
\begin{eqnarray}
\fl\mathring{f}(r,z,v_1,v_2,b)=A\big((r^{\max}-r)_+(r-r^{\min})_+(z^{\max}-z)_+(z-z^{\min})_+\big)^2\nonumber\\
\;\;\;\;\big((v_1^{\max}-v_1)_+(v_1-v_1^{\min})_+(v_2^{\max}-v_2)_+(v_2-v_2^{\min})_+\big)^2\nonumber\\
\;\;\;\;\big((b^{\max}-b)_+ (b-b^{\min})_+\big)^2,
\label{f-initial-1}
\end{eqnarray}
%\begin{eqnarray}
%\fl\mathring{f}(r,z,v_1,v_2,b)=A\big((r^{\max}-r)_+(r-r^{\min})_+(z^{\max}-z)_+(z-z^{\min})_+\big)^2\nonumber\\
%\big((v_1^{\max}-v_1)_+(v_1-v_1^{\min})_+(v_2^{\max}-v_2)_+(v_2-v_2^{\min})_+\big)^2\nonumber\\
%\big((b^{\max}-b)_+(b-b^{\mathrm{gapmax}})_+ + (b-b^{\min})_+(b^{\mathrm{gapmin}}-b)_+\big)^2,
%\label{f-initial-1}
%\end{eqnarray}
where $(x)_+=x$ if $x\geq 0$ and $(x)_+=0$ if $x<0$.
The parameter values $r^{\min}, r^{\max}, z^{\min}, z^{\max}, v_1^{\min}, v_1^{\max}, v_2^{\min}, v_2^{\max}, b^{\min}$ and  $b^{\max}$ determine the support of $\mathring{f}$. 
We call the parameter $A>0$ the amplitude. 
Furthermore, we always choose $z^{\min}=-z^{\max}<0$ and $b^{\min}=-b^{\max}$ to ensure that the total angular momentum is zero, cf. Equation (B.23) in \cite{Ames2021}, since our code requires that the total angular momentum vanishes as discussed above. 

The amplitude $A$ is chosen such that the ADM mass $M$ of the initial configuration, cf. Equation (B.18) in \cite{Ames2021}, is roughly $1$. This turns out to correspond to supercritical data so that collapse occurs. The time and length scales are then naturally given in terms of $M$. Although our initial configurations are not perfect ellipsoids they are highly elongated and we therefore associate the initial data with an eccentricity $e$ which we define to be 
$$e=\sqrt{1-\frac{r_{max}^2}{z_{max}^2}}.$$ 
This is in analogy with the eccentricity of an ellipsoid with equatorial radius $r_{max}$ and polar radius $z_{max}$.

Due to the large number of phase-space dimensions there is a vast freedom to vary the initial data.
First of all we limit ourselves to the form of the initial data given by (\ref{f-initial-1}) and for this choice we only consider a few cases. We focus on the highly prolate case to challenge the Hoop conjecture, and we always choose $r_{max}=0.5$ and $z_{max}=4$. This gives eccentricity $e=0.992$ which can be compared to the maximum eccentricity in \cite{East2019} which is $e=0.95$ and to the maximal eccentricity in \cite{Shapiro1991,Yoo2017} which is $e=0.9$. A direct comparison is however not possible since the shape of the initial density is different in our work compared to \cite{Shapiro1991,Yoo2017,East2019}. For the momentum variables we distinguish between two cases. Either we choose the parameters for the momentum variables $v_1$ and $v_2$ such that the initial data is time symmetric, i.e. the current vanishes, or we shoot the particles
inward initially to investigate if this has an impact on the collapse scenario. 
Naively, the latter case would be a more severe challenge to weak cosmic censorship. 

In addition to the initial data for the density function we also choose initial data for the fields. We choose trivial data as in \cite{Ames2021}, in particular, we do not include any gravitational wave degrees of freedom ``by hand". Furthermore, we have to specify the size of the numerical domain and we need to make it sufficiently large so that the boundary conditions are satisfied to reasonable tolerance. This issue is not present in spherical symmetry since in that case the Schwarzschild solution determines the geometry outside the support of the matter whereas in the present case the solution is not explicitly known outside the support of the matter. Hence, in spherical symmetry it is sufficient that the numerical grid covers the support of the matter whereas we need a rather large domain even if matter initially only occupies a small part of the domain. The size of our numerical domain is $(r,z)\in [0,15]\times [-30,30]$. 
 
We evolve the following families of initial data: 

\begin{table}[h!]
\begin{center}
  \def\arraystretch{1.5}
  \begin{tabular}{ |c|c|c|c|c|c|} 
    \hline
    ID Set & $r$ & $z$ & $v_1$ & $v_2$ & $b$ \\   
    \hhline{|=|=|=|=|=|=|} 
    ID1 
    & $[0.05,0.5]$ & $[-4.0,4.0]$ 
    & $[-0.2, 0.2]$ & $[-0.2, 0.2]$ 
    & $[-0.4, 0.4]$ \\     
    \hline
    ID2 
    & $[0.05,0.5]$ & $[-4.0,4.0]$ 
    & $[-0.2, 0.2]$ & $[-0.2, 0.2]$ 
    & $[-0.2, 0.2]$ \\    
    \hline 
%    ID3 
%    & $[0.0,0.5]$ & $[-4.0,4.0]$ 
%    & $[-0.2, 0.2]$ & $[-0.2, 0.2]$ 
%    & $[-0.4, -0.2] \cup [0.2, 0.4]$ \\ 
%    \hline
    ID3
    & $[0.05,0.5]$ & $[-4.0,4.0]$ 
    & $[-0.4, 0.0]$ & $[-0.2, 0.2]$ 
    & $[-0.4, 0.4]$ \\
    \hline
%    ID5
%    & $[0,0]$ & $[0,0]$ 
%    & $[0, 0]$ & $[0, 0]$ 
%    & $[0, 0] \cup [0, 0]$ \\
%    \hline
  \end{tabular}
  \caption{\label{t.initial_data}The parameter values appearing in (\ref{f-initial-1}), except for the amplitude $A$, are given in the table. For instance $v_1^{\min}=-0.2$ and $v_1^{\max}=0.2$ for family ID1. 
  }
  \end{center}
\end{table}
The data sets ID1 and ID2 are time symmetric whereas in the case ID3 the particles are shot inwards initially. The difference between ID1 and ID2 concerns the range of the angular momentum which is larger in the former case. 
We also remark that we have chosen $r_{\min}=0.05$ instead of $r_{\min}=0$. This choice slightly improves the momentum constraints. 
In view of the initial density function (\ref{f-initial-1}) we note that the amplitude of $\mathring{f}$ vanishes at $r=r_{\min}$ so that even if particles were placed at $r=r_{\min}=0$ they would not be noticeable initially due to the choice of $\mathring{f}$. 

In \ref{numerical_accuracy} we present some convergence results by varying the resolution in phase space and in Section \ref{S8} we present the results of our simulations.

\section{Numerical results}\label{S8}
Qualitatively the properties of the evolved solutions of the different data in Table 1 are quite similar. We give a detailed description of the properties corresponding to these data below. 
The evolution of the energy density for each data is shown in Figure \ref{f.time_series_plots}. 

An apparent horizon is found using the method outlined in \cite[Sect. 3.3]{Ames2021}. Figure \ref{f.horizon_shape_allID} depicts the shape of the horizon roughly when it forms for each initial data set.
We stop the simulation shortly after an apparent horizon has been found since we then experience increasing violations of the constraints and of the mass conservation.
In each case the horizon is mildly prolate.

\begin{figure}[htb]
  \centering
  \includegraphics[height=0.75\textwidth]{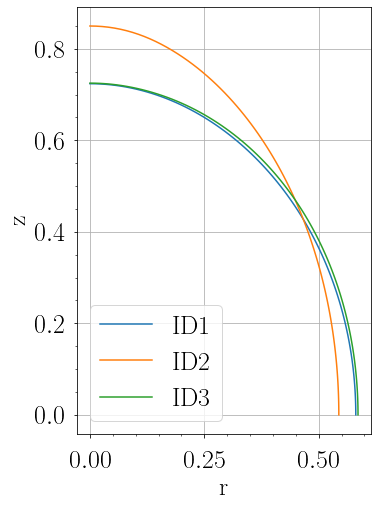}
  \caption{
    \label{f.horizon_shape_allID} 
    Horizon curves in the (r,z)-plane at the time of apparent horizon formation for each initial data set. 
  }
\end{figure}

Polar and equitorial circumference information is provided in Table \ref{t.horizon_information}, along with horizon formation time and mass.
By computing the horizon mass by the strategy outlined in Section \ref{S45} we can test the Hoop conjecture by computing $\mathcal{\kappa}_p$ and $\mathcal{\kappa}_e$ given by (\ref{kappa}). 

\begin{table}[h!]
  \begin{center}
    \def\arraystretch{1.5}
    \begin{tabular}{ |c|c|c|c|c|c|c|} 
      \hline
      ID Set & $t_H$ & $M_H$ & 
      $\mathcal{C}_{H,p}$ & $\mathcal{C}_{H,e}$ &
      $\mathcal{\kappa}_p$ & $\mathcal{\kappa}_e$ \\   
      \hhline{|=|=|=|=|=|=|=|} 
      ID1 
      & $11.71$ & $1.0$ & $12.57$ & $10.86$ & $1.00$ & $0.87$  \\     
      \hline
      ID2 
      & $9.80$ & $0.99$ & $13.83$ & $10.53$ & $1.12$ & $0.85$ \\
      \hline 
      ID3
      & $11.30$ & $0.98$ & $12.27$ & $10.58$ & $1.00$ & $0.86$ \\
      \hline
    \end{tabular}
    \caption{
      \label{t.horizon_information}
      The (numerical) time of horizon formation $t_H$, the horizom mass $M_H$, polar and equitorial circumferences $\mathcal{C}_{H,p}$ and $\mathcal{C}_{H,e}$, and quantities $\mathcal{\kappa}_p$ and $\mathcal{\kappa}_e$ as defined in Section (\ref{S45}) for each family of initial data.
    }
    \end{center}
  \end{table}

We notice that the value of $\mathcal{\kappa}_p$ in each case is very close to one (for ID1 $\mathcal{\kappa}_p=1.00$, and for ID3 $\mathcal{\kappa}_p=1.00$).
Interestingly, this feature also holds in a few other cases (not reported here), which leads us to believe that may be a general feature. 
However, for initial data ID2 $\mathcal{\kappa}_p=1.12$. 
Since the errors are slightly bigger for this initial data set (note the large curvature in Figure \ref{f.curvature_scalars}) it is possible the discrepancy is numerical error and that the relation $\mathcal{\kappa}_p \equiv 1$ at the time of horizon formation and for sufficiently prolate initial data still holds. 
We are curious to investigate this feature more carefully in a larger variety of initial data families.
In any case, our result that $\mathcal{\kappa}_p$ is above 1 by 12\% is commensurate with the results in \cite{East2019} where $\mathcal{\kappa}_p$ and $\mathcal{\kappa}_e$ vary between 0.75 and 1.25.

Let us return to discuss some other properties of the solutions. 
In Figure \ref{f.energy_density_alphamin} the maximum value of the energy density $\rho_H$ is shown (left panel). 
There is a rapid increase before the apparent horizon forms and it continues to increase after the formation (not shown) which agrees with the expectation in gravitational collapse. 
Figure \ref{f.energy_density_alphamin} (right panel) shows how the minimum value of the metric field $\alpha$ (the lapse) evolves, and as expected it decreases and becomes very small when the apparent horizon forms. 

\begin{figure}[htb]
  \centering
  \begin{subfigure}[b]{0.48\linewidth}
    \includegraphics[height=0.65\textwidth]{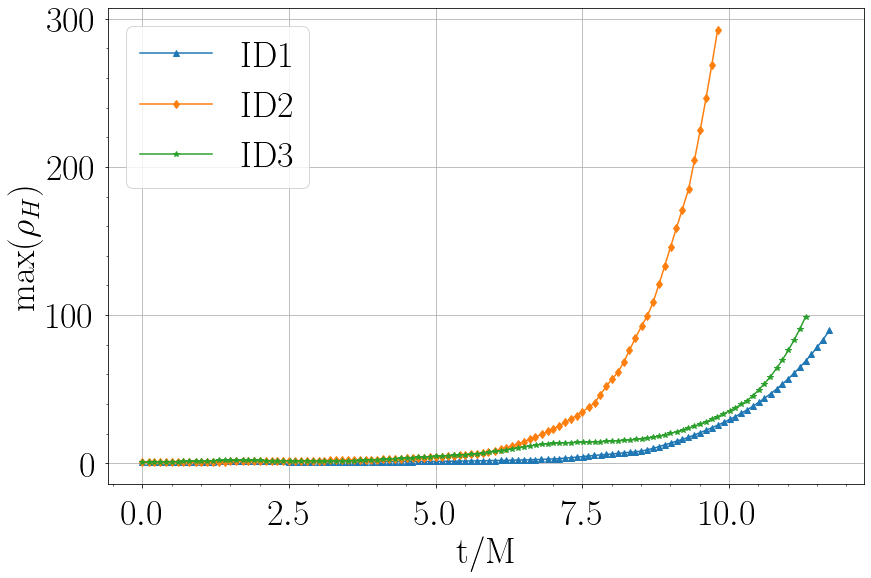}
  \end{subfigure}\quad
  \begin{subfigure}[b]{0.48\linewidth}
    \includegraphics[height=0.65\textwidth]{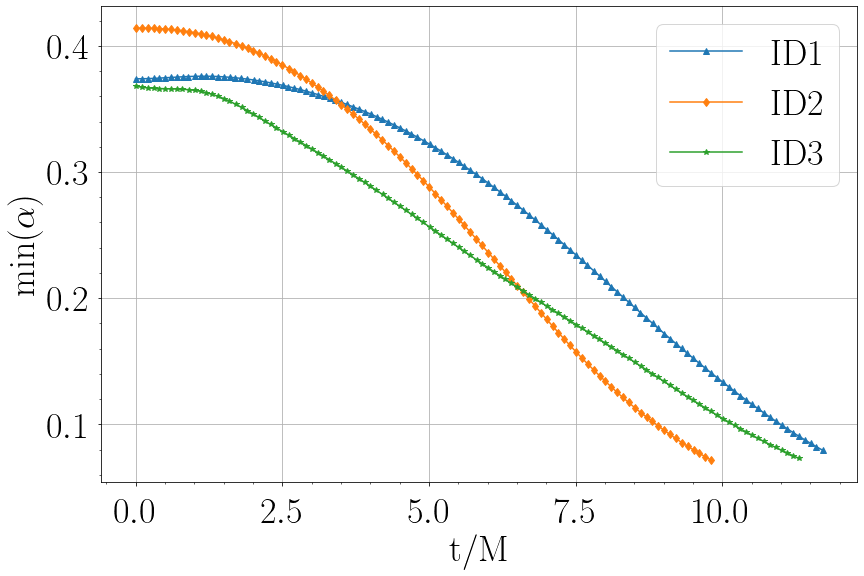}
  \end{subfigure}
  \caption{
    \label{f.energy_density_alphamin} 
    Left panel: The maximum of the energy density $\rho_H$.
    Right panel: The minimum value of $\alpha$.
    These quantities support the conclusion that no singularities form during the evolution before the formation of an apparent horizon.  
  }
\end{figure}

Figure \ref{f.curvature_scalars} shows the maximum of the Kretschmann scalar (left panel) and scalar curvature (right panel).
The evolution of each quantity exhibits a period of rapid increase (particularly for ID2), but eventually stabilizes (in the case of ID3 even decreases), before an apparent horizon eventually forms.
This bounded behavior of the curvature scalars supports the conclusion that no singularities form before the apparent horizon.

\begin{figure}[htb]
  \centering
  \begin{subfigure}[b]{0.48\linewidth}
    \includegraphics[height=0.65\textwidth]{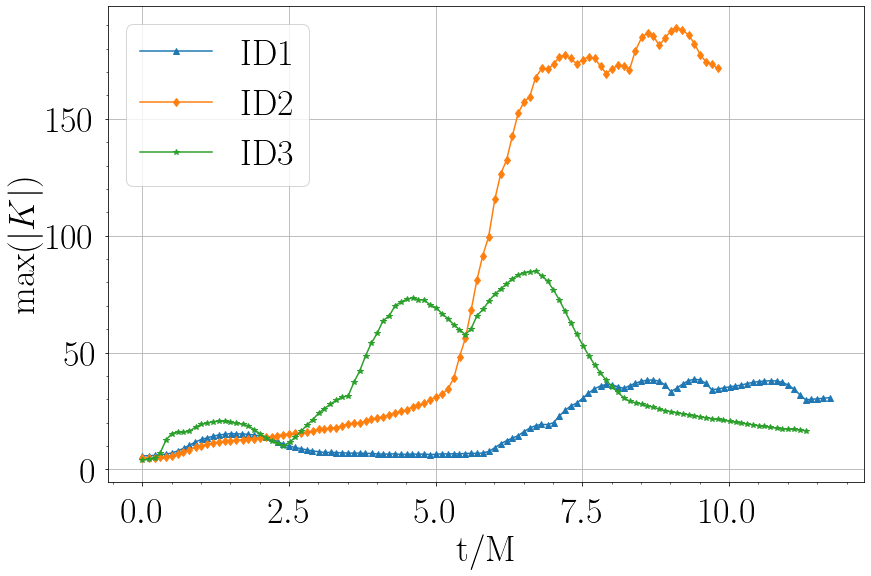}
  \end{subfigure}\quad
  \begin{subfigure}[b]{0.48\linewidth}
    \includegraphics[height=0.65\textwidth]{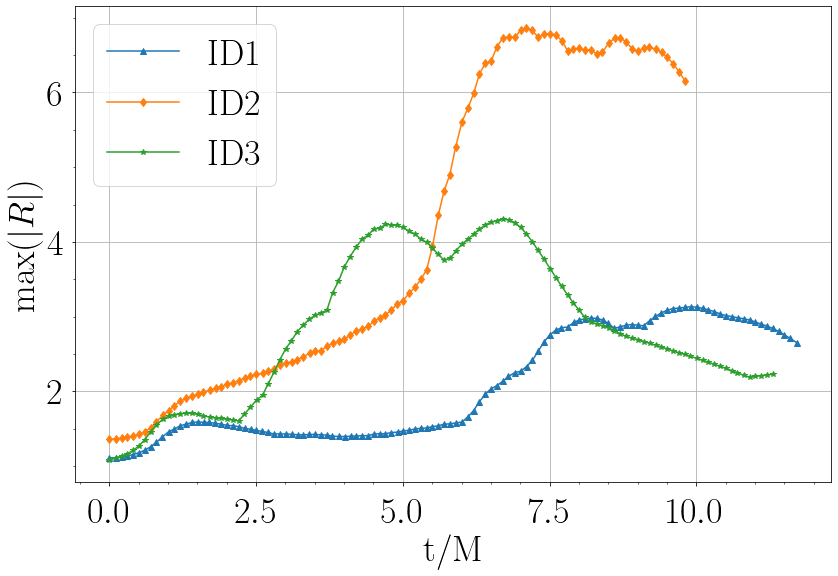}
  \end{subfigure}
  \caption{
    \label{f.curvature_scalars} 
    Left panel: Maximum value of the Kretschmann scalar.
    Right panel: Maximum value of the scalar curvature.
  }
\end{figure}

The behavior of the innermost particles (those closest to the axis) is shown in Figure \ref{f.rpmin_IDall_ar16_9}, which shows the minimum value of the radial particle position.
Initially the innermost particles move inward, then turn around and move outward for some time due to the angular momentum that each particle carries, before finally turning back inward again in the phase leading to collapse.
One notes that the minimum value is roughly $0.07$ and not $0.05$ as in Table \ref{t.initial_data}. 
This is because, due to the initial data profile (\ref{f-initial-1}), particles at the boundary of the support have zero amplitude and are removed before the evolution.

% The quantities associated with the behaviour of the particles are displayed in Figures \ref{f.rpmin_v1p_support}-\ref{f.v2p_bp_support}. 
% % In Figure \ref{rmax} the maximum value, ranging over all particles, of the cylindrical radius is depicted. It is denoted by $\textrm{rpmax}$. The notation for the other particle quantities is analogous. We note in Figure \ref{rmax} that initially all particles are situated within cylindrical radius $r\leq 0.5$, and at time $t_{H}$ the outermost particles are further out than $r=5.0$. 
% The minimum value of the radial particle position is shown in the left panel of Figure \ref{rpmin_v1p_support}. 
% Initially the innermost particles move inward, then turn around and move outward for some time due to the angular momentum that each particle carries, before finally turning back inward again in the phase leading to collapse. 
% Corresponding figures for the momentum variables $v_1$ and $v_2$ are given in Figures \ref{v1max}-\ref{v2min}. 
% By comparing Figures \ref{zmax} and \ref{zmin} with Figures \ref{v2max} and \ref{v2min} we note that the maximum value $\textrm{v2pmax}$ and minimum value $\textrm{v2pmin}$ are carried by particles moving toward the equator.

\begin{figure}[h!]
  \centering
  \includegraphics[width=0.75\textwidth]{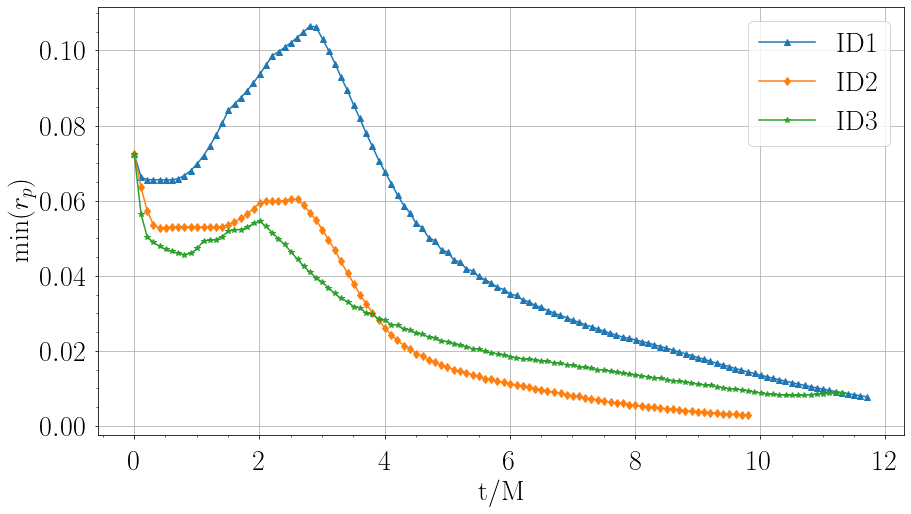}
  \caption{ 
    \label{f.rpmin_IDall_ar16_9}
    The minimum $r$-coordinate of a particle for each initial data set.
  }
\end{figure}

% \begin{figure}[htb]
%   \centering
%   \begin{subfigure}[b]{0.48\linewidth}
%     \includegraphics[height=0.65\textwidth]{figures/rpmin_IDall.png}
%   \end{subfigure}\quad
%   \begin{subfigure}[b]{0.48\linewidth}
%     \includegraphics[height=0.65\textwidth]{figures/v1p_support_IDall.png}
%   \end{subfigure}
%   \caption{
%     \label{f.rpmin_v1p_support} 
%     Left panel: The minimum $r$-coordinate of a particle for each initial data set.
%     Right panel: The envelope of support for each initial data set in the $v_1$-momentum. 
%   }
% \end{figure}

% \begin{figure}[htb]
%   \centering
%   \begin{subfigure}[b]{0.48\linewidth}
%     \includegraphics[height=0.65\textwidth]{figures/v2p_support_IDall.png}
%   \end{subfigure}\quad
%   \begin{subfigure}[b]{0.48\linewidth}
%     \includegraphics[height=0.65\textwidth]{figures/bp_support_IDall.png}
%   \end{subfigure}
%   \caption{
%     \label{f.v2p_bp_support} 
%     The envelope of support for each initial data set in the $v_2$-momentum (left panel), and in the $b$-momentum (right panel).
%     Note that points for ID1 (in blue) are mostly behind those corresponding to ID3 (in green). 
%   }
% \end{figure}

As is seen in Figure \ref{f.time_series_plots}, where snapshots of the energy density $\rho_H$ are displayed at different times in the evolution, the matter configuration changes from a highly prolate shape to a mildly prolate shape at the time when an apparent horizon forms.
Note that the colour scale is logarithmic, and the ``atmosphere" extending outside the ``core" of the matter is thus much thinner than it appears to be.  
Hence, the matter configuration can be regarded as consisting of a core and of a thin atmosphere where the core eventually collapses and forms a black hole. 
The atmosphere consists of particles that are ejected outwards from the core of the matter. 
It is clear from the pictures that there is no well defined boundary of the core as discussed in Section \ref{S45}. 
That the atmosphere is thin is confirmed by the mass of the black hole which is less than but close to the initial mass. 

\begin{figure}[htb]
  \centering
  \begin{subfigure}[b]{\linewidth}
    \includegraphics[width=\textwidth]{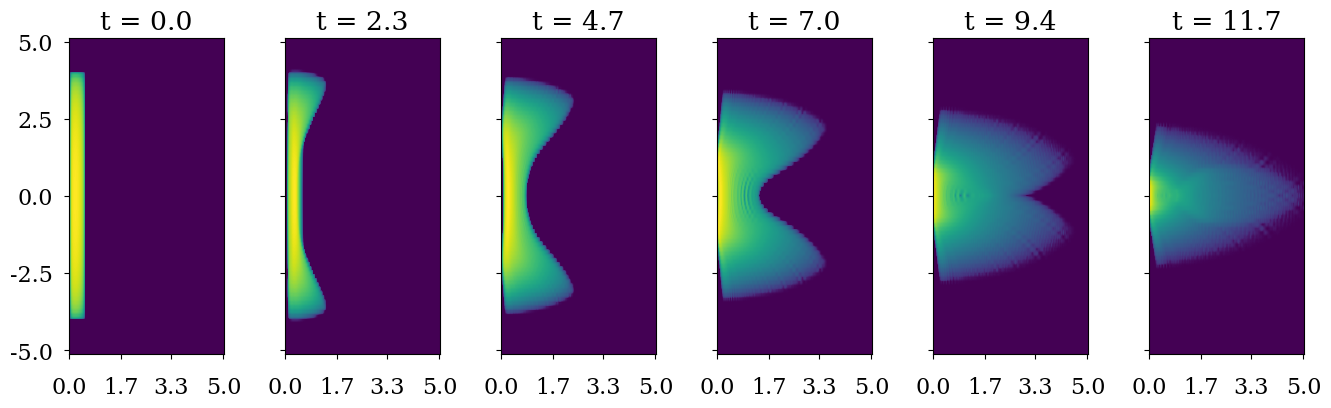}
  \end{subfigure}\\

  \vspace{1ex}

  \begin{subfigure}[b]{\linewidth}
    \includegraphics[width=\textwidth]{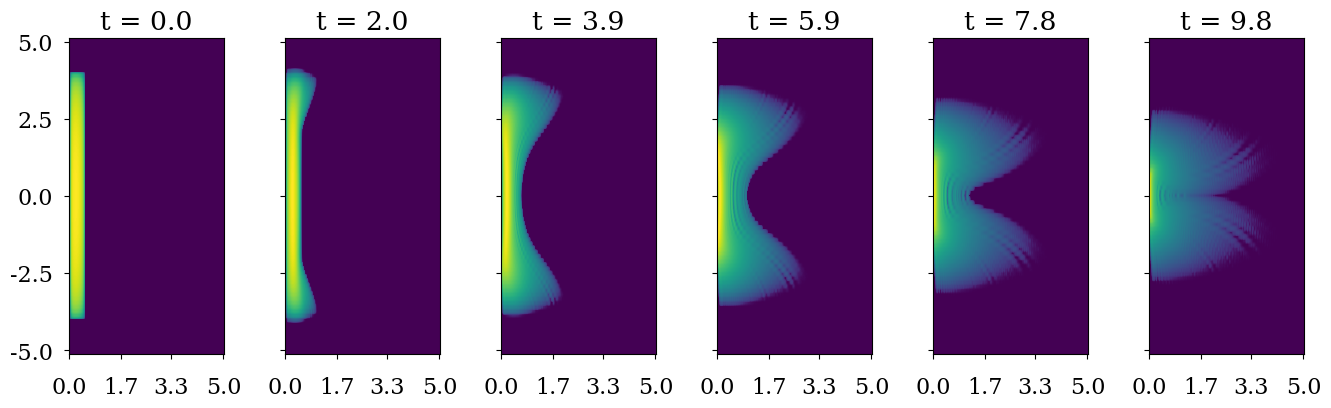}
  \end{subfigure}\\

  \vspace{1ex}

  \begin{subfigure}[b]{\linewidth}
    \includegraphics[width=\textwidth]{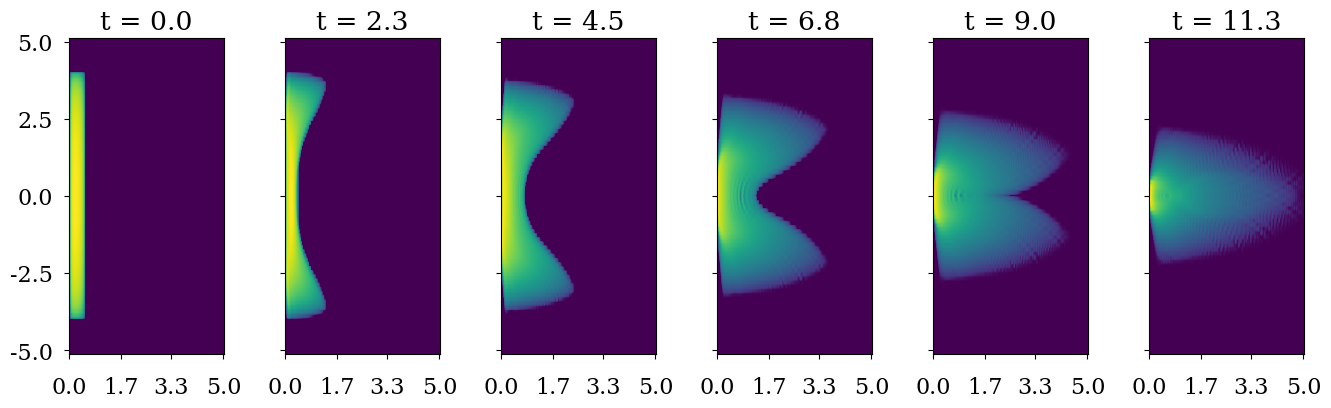}
  \end{subfigure}
  \caption{
    \label{f.time_series_plots} 
    Matter density $\rho_H$ for each initial data set (top: ID1, middle: ID2, bottom: ID3). 
    The color scale is logarithmic above $10^{-6}$, and differs for each initial data set. 
    See Figure \ref{f.energy_density_alphamin} for an estimate of the peak values. 
    Times are in units of $M$.
  }
\end{figure}

Hence we find that in the evolution of initial data which could be a potential threat to weak cosmic censorship, matter drastically changes shape until an apparent horizon forms which is only mildly prolate and which satisfies the constraints of the Hoop conjecture. This behaviour is similar to what we found in our previous work \cite{Ames2021} where we challenged weak cosmic censorship by trying to collapse initial data with total angular momentum $|J|>M^2$. Note that a Kerr solution with $|J|>M^2$ has a naked singularity. We found that in the evolution particles were ejected in an oscillatory phase until the total angular momentum of the core of the matter, $J_{core}$, satisfied the reverse inequality $|J_{core}|<M^2$. Once this happened an apparent horizon formed and the core collapsed to a black hole saving weak cosmic censorship, cf. Section 4.3.3 in \cite{Ames2021}. 

In conclusion we have found strong numerical support for both the weak cosmic censorship conjecture and the Hoop conjecture for solutions of the (regular) axially symmetric Einstein-Vlasov system.

\section*{Acknowledgments}

This research was partially supported through the programme ``Research in Pairs" by the Mathematisches Forschungsinstitut Oberwolfach in 2021. 
Computations were performed on resources at Chalmers Centre for Computational Science and Engineering (C3SE) provided by the Swedish National Infrastructure for Computing (SNIC). 

%%%%%%%%%%%%%%%%%%%%%%%%%%%%%%%%%%%%%%%%%%%%%%%%%%%%%%%%%%%%%%%%%%%%%%%%%
%%%%%%%%%%%%%%%%%%%%%%%%%%%%%%%%%%%%%%%%%%%%%%%%%%%%%%%%%%%%%%%%%%%%%%%%%
%%%%%%%%%%%%%%%%%%%%%%%%%%%%%%%%%%%%%%%%%%%%%%%%%%%%%%%%%%%%%%%%%%%%%%%%%

\appendix

\section{Tests of numerical accuracy}
\label{numerical_accuracy}
In addition to the parameters in Table 1 we also make choices of the resolution in phase space. As in \cite{Ames2021} we choose the parameters $N_r, N_z,N_1^v,N_2^v$ and $N_3^v$, where the latter corresponds to the momentum variable $b$ in this work. The number of particles that we evolve in the code depends on the resolution. In a high-resolution simulation we evolve a few million particles which is more than, but comparable to, the number of particles evolved in \cite{East2019,Yoo2017}. The number of particles evolved in \cite{Shapiro1991} was 6000. 

\begin{table}[h!]
  \begin{center}
    \begin{tabular}{ |c|c|c|c|c|c|} 
      \hline
      Resolution & $N_r$ & $N_z$ & $N_1^v$ & $N_2^v$ & $N_3^v$ \\ 
      \hhline{|=|=|=|=|=|=|} 
      R1
      & $350$ & $225$ 
      & $8$ & $8$ 
      & $8$ \\         
      \hline 
      R2
      & $500$ & $320$ 
      & $12$ & $12$ 
      & $12$ \\     
      \hline
      R3 
      & $700$ & $450$ 
      & $16$ & $16$ 
      & $16$ \\ 
      \hline
        \end{tabular}
    \caption{\label{t.resolution}}
    \end{center}
  \end{table}

The ADM mass $M$ and the total angular momentum $J$ are conserved quantities and it is natural to check the accuracy of the code by monitoring these quantities. Since the initial data in all cases studied have total angular momentum equal to zero this quantity does not reveal much information and we leave it out. (Typically $|J|<10^{-17}$ in our simulations.) As described in \cite{Ames2021} the momentum constraints are not solved in the evolution scheme and it is essential to monitor the residuals of the constraints to check the accuracy of the code. 

\begin{figure}[h!]
  \centering
  \includegraphics[width=0.75\textwidth]{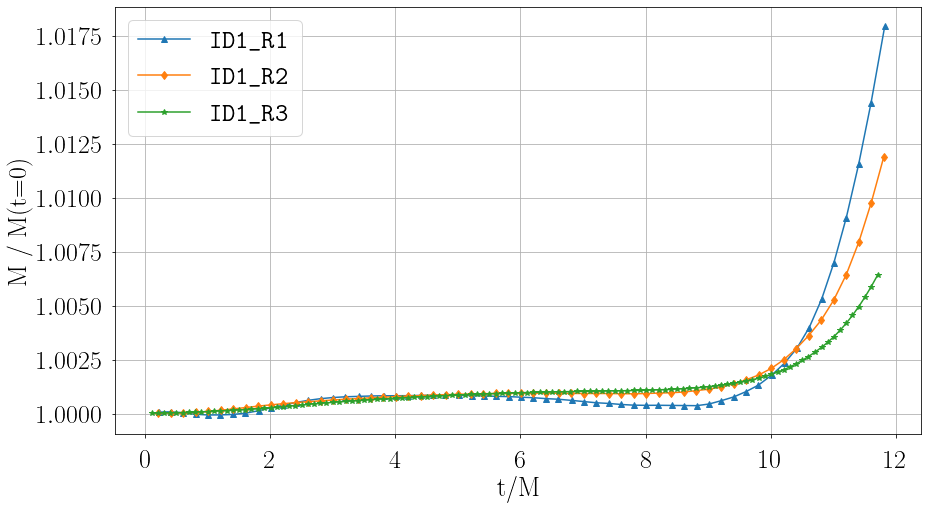}
  \caption{ 
    \label{ADMmassM}
    ADM mass as function of time for ID1 with different resolutions. Formation of an apparent horizon occurs at $t\approx 11.71$.
  }
\end{figure}

We have done a convergence test for the initial data class ID1 in Table 1 with three different resolutions: R1, R2 and R3. In Table 2 the choices of the parameters $N_r, N_z, N_1^v, N_2^v$ and $N_3^v$ are given for the different cases. The ADM masses for the different resolutions are depicted in Figure \ref{ADMmassM}. Up to the time $t_{H}$ when an apparent horizon forms, which in these simulations is roughly at $t_{H} \approx 11.7$, the ADM mass grows slightly but within two percent. In the case of highest resolution, i.e. resolution R3, the growth is less than $1$ percent up to $t=t_{H}$. The residuals for the normalized momentum constraints are depicted in Figure \ref{f.constraint_residuals}. 
In the case of the highest resolution the error stays below $10\%$ up to time $t_{H}$. 
This is comparable to the convergence results presented in \cite{Yoo2017,East2019}. 

\begin{figure}[htb]
  \centering
  \begin{subfigure}[b]{0.75\linewidth}
    \includegraphics[width=\textwidth]{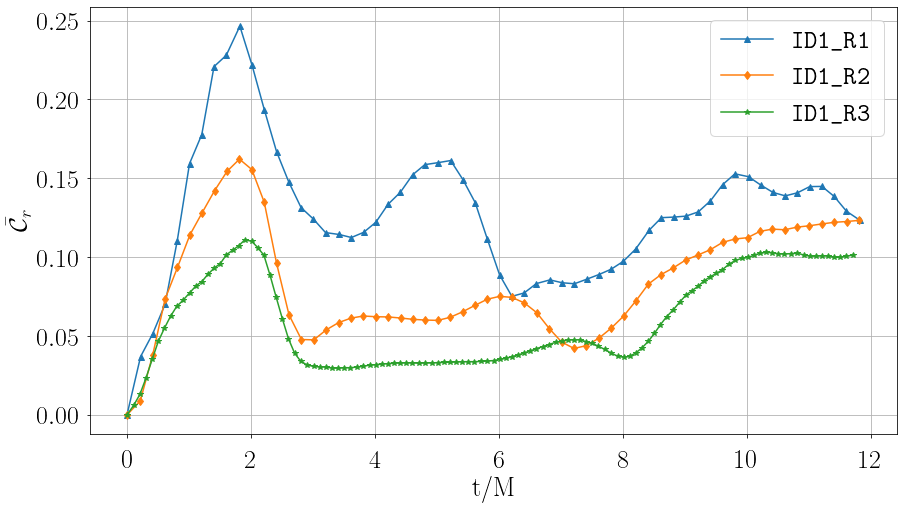}
  \end{subfigure}\quad
  \begin{subfigure}[b]{0.75\linewidth}
    \includegraphics[width=\textwidth]{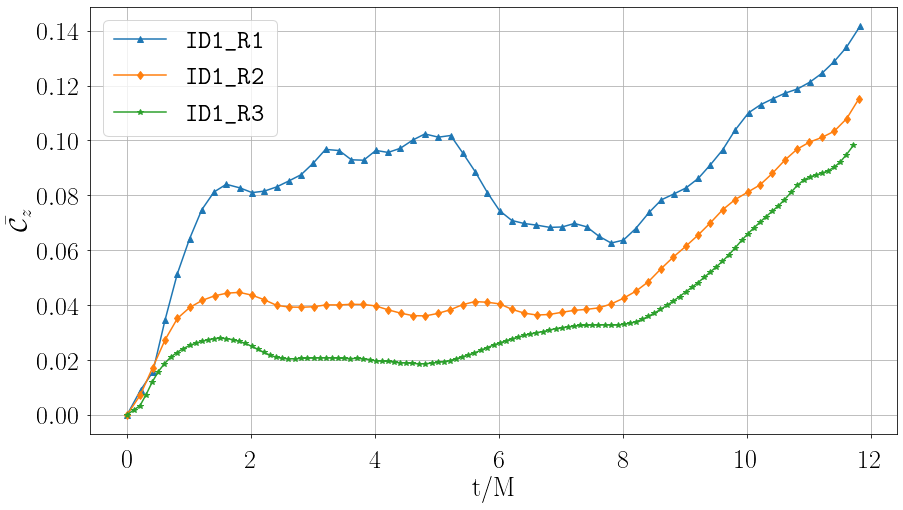}
  \end{subfigure}
  \caption{
    \label{f.constraint_residuals} 
    Normalized residuals for the unsolved momentum constraint equations. 
    In the top panel $\overline{\mathcal C}_r$, and in bottom panel $\overline{\mathcal C}_z$, see \cite[equation (77)]{Ames2021} for definitions.
    Low (blue), mid (orange), and high (green) resolution, as defined in Table \ref{t.resolution}, are shown from top to bottom.
  }
\end{figure}

%%%%%%%%%%%%%%%%%%%%%%%%%%%%%%%%%%%%%%%%%%%%%%%%%%%%%%%%%%%%%%%%%%%%%%%%%
%%%%%%%%%%%%%%%%%%%%%%%%%%%%%%%%%%%%%%%%%%%%%%%%%%%%%%%%%%%%%%%%%%%%%%%%%
%%%%%%%%%%%%%%%%%%%%%%%%%%%%%%%%%%%%%%%%%%%%%%%%%%%%%%%%%%%%%%%%%%%%%%%%%

\section*{References}

\bibliographystyle{iopart-num-long}
\bibliography{references}

\end{document}